\documentclass[12pt,a4paper]{article}

\usepackage{times}

\usepackage{graphicx}

\usepackage[table]{xcolor}
\usepackage{amsmath}
\usepackage{multirow}
\usepackage{bbding}

\usepackage[textheight=24.7cm,textwidth=18.3cm]{geometry}

\usepackage[backend=biber,style=nature]{biblatex}
\AtEveryBibitem{%
\ifentrytype{article}{
    \clearfield{url}%
    \clearfield{issue}%
    \clearfield{month}%
    \clearfield{URL}%
    \clearfield{eprint}
    \clearfield{issn}
}{}
}

\newcommand{\mean}[1]{\left\langle#1\right\rangle}

\title{One-hundred step measurement-based quantum computation
multiplexed in the time domain with 25 MHz clock frequency}

\author
{Warit Asavanant,$^{1,\dagger,\dagger \dagger}$ Baramee Charoensombutamon,$^{1,\dagger}$ Shota Yokoyama,$^{2}$ \\  Takeru Ebihara,$^{1}$ Tomohiro Nakamura,$^{1}$ Rafael N. Alexander,$^{3}$\\ Mamoru Endo,$^{1}$ Jun-ichi Yoshikawa,$^{1}$ Nicolas C. Menicucci,$^{4}$\\ Hidehiro Yonezawa,$^{2}$ Akira Furusawa$^{1,\ast}$\\
\\
\normalsize{$^{1}$Department of Applied Physics, School of Engineering, The University of Tokyo,}\\
\normalsize{7-3-1 Hongo, Bunkyo-ku, Tokyo 113-8656, Japan}\\
\normalsize{$^{2}$Centre for Quantum Computation and Communication Technology,}\\
\normalsize{School of Engineering and Information Technology, University of New South Wales,}\\
\normalsize{Canberra, ACT 2600, Australia}\\
\normalsize{$^{3}$ Center for Quantum Information and Control, Department of Physics and Astronomy,}\\
\normalsize{University of New Mexico, Albuquerque, NM 87131, USA}\\
\normalsize{$^{4}$Centre for Quantum Computation and Communication Technology,}\\
\normalsize{School of Science, RMIT University, Melbourne, Victoria 3001, Australia}\\
\\
\normalsize{$^{\dagger}$ These authors contributed equally to this work}\\
\normalsize{$^\ast$To whom correspondence should be addressed; E-mail:  akiraf@ap.t.u-tokyo.ac.jp.}\\
\normalsize{$^{\dagger\dagger}$ E-mail: warit@alice.t.u-tokyo.ac.jp.}\\
}
\date{}

\begin{document} 

% Double-space the manuscript.

\baselineskip24pt

% Make the title.

\maketitle 

% Place your abstract within the special {sciabstract} environment.

{\bf Quantum computers are known to provide algorithmic speed ups over their classical counterparts\supercite{nielsen00}. In recent years, approaches based on various physical systems---superconducting qubits\supercite{doi:10.1063/1.5089550}, ion-trap systems\supercite{doi:10.1063/1.5088164}, and photonic systems\supercite{doi:10.1063/1.5100160,doi:10.1063/1.5115814}---have been extensively explored. However, constructing devices at scale required for real-world applications is no trivial task. Among the various approaches, measurement-based quantum computation\supercite{PhysRevLett.86.5188,PhysRevLett.97.110501} (MBQC) multiplexed in time domain\supercite{PhysRevA.83.062314} is currently a promising method for addressing the need for scalability. MBQC requires two components: cluster states and programmable measurements. With time-domain multiplexing, the former has been realized on an ultra-large-scale\supercite{Yokoyama2013,doi:10.1063/1.4962732,Asavanant373,Larsen369}. The latter, however, has remained unrealized, leaving the large-scale cluster states unused. In this work, we make such a measurement system and use it to demonstrate basic quantum operations multiplexed in the time-domain with 25 MHz clock frequency. This programmable measurement system is a key piece for MBQC in time domain. We directly verify transformations of the input states and their nonclassicalities for single-step quantum operations and also observe multi-step quantum operations up to one hundred steps---the largest number ever observed in CV systems. Furthermore, with sufficient squeezing and bosonic qubits, cluster state and programmable measurement system in this work serve as the central components of universal fault-tolerant MBQC\supercite{PhysRevA.64.012310,PhysRevLett.112.120504,PhysRevX.8.021054,PhysRevA.100.010301}. Our work establishes a prototype for scalable quantum computation based on CV optical systems and demonstrates a new and promising road toward full-fledged quantum computation.}

The promise of quantum computing technology stems from a wide range of potential applications, and has spawned many alternative realizations based on different physical systems\supercite{doi:10.1063/1.5089550,doi:10.1063/1.5088164,doi:10.1063/1.5100160,doi:10.1063/1.5115814}. Quantum supremacy using superconducting qubits\supercite{Arute2019} and Boson sampling\supercite{PhysRevLett.123.250503} has recently been realized. While these achievements mark important milestones in the journey towards quantum computation, these experiments are still a far cry from useful quantum computations. A fault-tolerant quantum computer that can handle arbitrary large-scale multi-input and multi-step computation is essential. With matter-based qubits, achieving quantum computation on a useful scale involves preparing and interfacing a large number of qubits, a task that would require a technological leap to reach the scale necessary for real-world applications.

\begin{figure*}[p]
\centering
\includegraphics{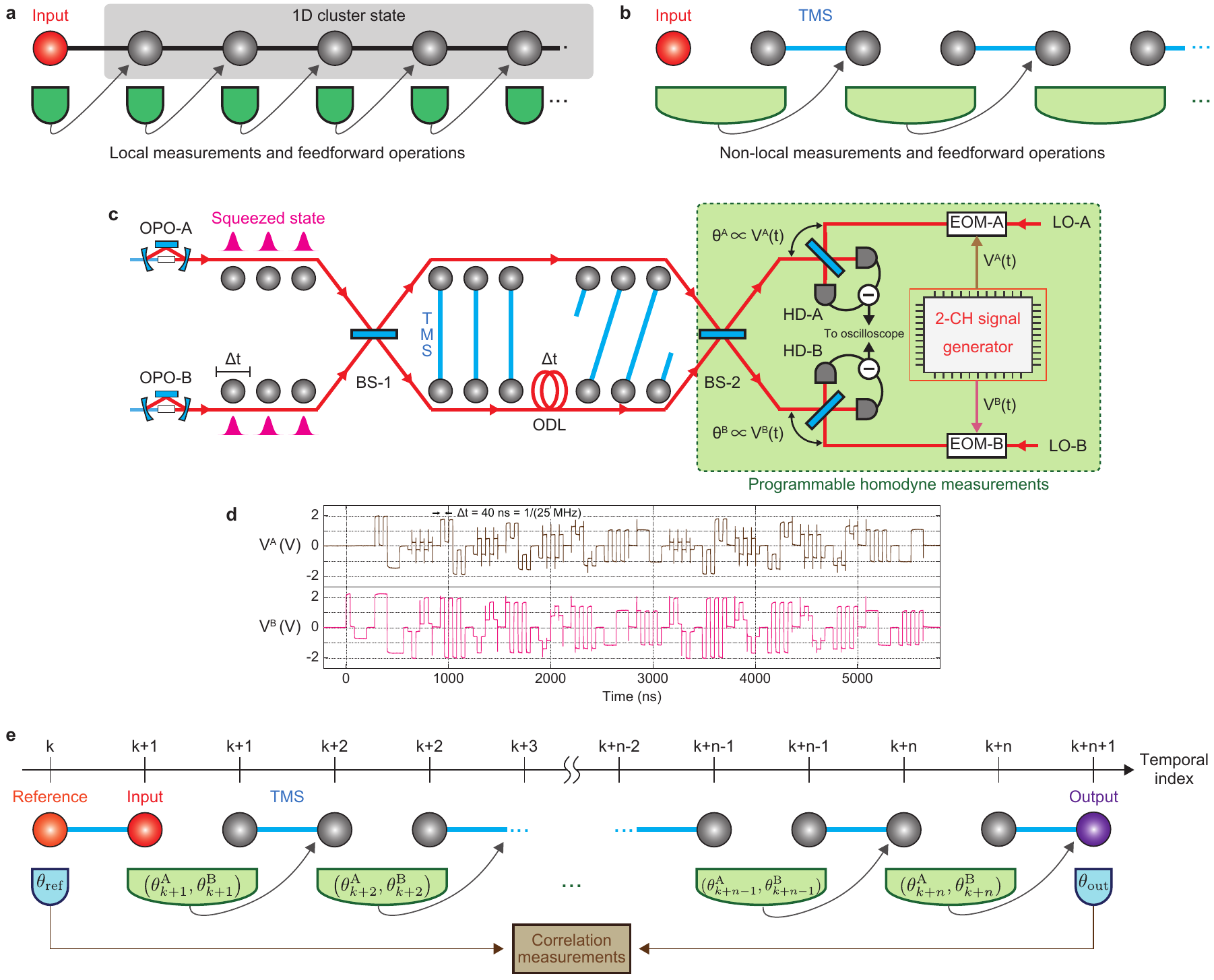}
\caption{\textbf{MBQC in time domain and its verification.} MBQC using \textbf{a,} local measurements on 1D cluster states and \textbf{b,} non-local measurements on two-mode-squeezed states. The term \lq\lq{}non-local\rq\rq{} measurement here refers to local homodyne measurements on each mode after a beamsplitter interaction. \textbf{c,} Experimental system for MBQC multiplexed in time domain. TMS, two-mode-squeezed state; OPO, optical parametric oscillator; BS, beamsplitter; ODL, optical delay line; $\Delta t$, time width of the wave packet; HD, homodyne detector; LO, local oscillator; EOM, electro-optic modulator. The phases of each LO ($\theta^A$ and $\theta^B$) are proportional to $V^\textrm{A}(t)$ and $V^\textrm{B}(t)$, respectively. \textbf{d,} An example of the electric signals we can generate and use for switching measurement bases. The details of the circuitry are discussed in Method. \textbf{e,} Verification of $n$-step quantum operations using quantum entanglement. Measurement bases of the non-local measurements are determined by the bases $\theta^\textrm{A}$ and $\theta^\textrm{B}$ at HD-A and HD-B. Feedforward operations can be delayed and implemented numerically at the end for Gaussian operations. The blue detectors show the measurements of the reference mode and the output mode at bases $\theta_\textrm{ref}$ and $\theta_{\textrm{out}}$ (see Method for more details on this).}
\label{fig:TDMMBQC}
\end{figure*}

MBQC using optical modes multiplexed in the time domain\supercite{PhysRevA.83.062314} can overcome these obtacles. In MBQC, the requirements of qubit preparation and interfacing are replaced with a multi-qubit entangled resource, known as a cluster state\supercite{PhysRevLett.86.910}. After the cluster state is prepared, quantum operations are implemented via local (single) qubit measurements and feedforward operations based on the measurement results (Fig.\ \ref{fig:TDMMBQC}a). These features make MBQC an appealing experimental candidate as state preparations and measurements tend to be simpler than direct implementations of quantum gates. Generation of cluster states\supercite{Yokoyama2013,doi:10.1063/1.4962732,Asavanant373,Larsen369,Schwartz434,Reimer2019,PhysRevLett.122.110501,PhysRevLett.111.210501,PhysRevLett.112.120505,Walther2005} and the demonstration of small-scale MBQC\supercite{Reimer2019,PhysRevLett.111.210501,Walther2005,PhysRevLett.106.240504,Gao2011,PhysRevA.101.023809} in various systems have both already been achieved. To move from small-scale MBQC to large-scale MBQC, large-scale cluster states are required. In most approaches, however, the number of the experimental components scales directly with the size of the cluster states, making scaling-up difficult. In CV optical systems, by multiplexing temporally localized optical wave packet modes (CV equivalence to qubits) on the same beam, large-scale cluster states---10,000-mode\supercite{Yokoyama2013} and 1,000,000-mode\supercite{doi:10.1063/1.4962732} one-dimensional cluster states and universal resources for multi-input MBQC called two-dimensional cluster states\supercite{Asavanant373,Larsen369}---have been deterministically generated. These experimental results demonstrate the capability of time-domain-multiplexing (TDM) architecture in implementing a large-scale quantum computer.

With large-scale cluster states generated, a few additional ingredients are required to achieve a full-fledged quantum computer multiplexed in time domain: a programmable homodyne measurement system, a non-Gaussian element, and fault-tolerance. Cluster states form the backbone of this architecture. How we measure them, i.e. the choices of the measurement basis, determines the computation. Since modes are localized wave packets in the TDM method, high-speed switching and synchronicity, in addition to the programmability, are also necessary. Without the development of such systems, MBQC using TDM cluster states cannot be realized. Quantum computation with only CV cluster states and homodyne measurements are known to be classically simulatable\supercite{PhysRevLett.88.097904}; to achieve useful quantum computation, non-Gaussian elements must be added. There are various theoretical approaches to achieve this\supercite{PhysRevA.64.012310,PhysRevA.97.022329,PhysRevA.93.022301,PhysRevLett.123.200502} and non-Gaussian ancillary states with up to three photons have also been experimentally demonstrated\supercite{Yukawa:13}. Finally, to achieve fault-tolerant MBQC, we require the combination of sufficiently high quality cluster states and bosonic codes, of which the Gottesman-Kitaev-Preskill (GKP) qubit\supercite{PhysRevA.64.012310} is currently the most promising candidate\supercite{PhysRevLett.112.120504,PhysRevX.8.021054,PhysRevA.100.010301}. GKP qubits have been realized in microwave\supercite{2019arXiv190712487C} and ion-trapped systems\supercite{Fluhmann2019}, and various generation methods are being explored in optical systems\supercite{Hastrup:20,PhysRevA.97.022341,PhysRevA.101.032315}. 

In this work, we develop a programmable homodyne measurement system and demonstrate a prototype of a CV optical quantum computer using MBQC multiplexed in the time domain. Combining this system with our one-dimensional cluster state setup, we implement various quantum operations. We verify that quantum gates can be correctly programmed by the appropriate choice of measurement basis. Next, we also demonstrate the nonclassicality of the quantum operations by verifying that pre-shared entanglement persists between different mode even after quantum operations are applied to them. We verify that our system can implement arbitrary single-mode Gaussian operations (analogous to Clifford operations in qubits) and that the nonclassical features of the quantum states are preserved. Finally, we implement multi-step quantum operations and observe that our system has the stability and is capable of implementing MBQC up to at least one hundred steps. Not only is this the largest number of operations ever observed in CV systems, our work also demonstrates a working scalable and programmable CV optical quantum computer prototype. 

CV optical systems utilize electric fields---quadratures---as physical quantities in the quantum computation. The quadrature operators $\hat{x}$ and $\hat{p}$ associated with each temporal mode wave packet satisfy the following commutation relations: $[\hat{x}_{k},\hat{p}_{l}]=i\hbar\delta_{kl}$, where $k$ and $l$ are temporal indices of the wave packets. The basic measurement in CV optical systems is a homodyne measurement, which measures a linear combination of $\hat{x}$ and $\hat{p}$ for a given mode, i.e. $(\cos\theta)\hat{x}+(\sin\theta)\hat{p}$. We can express CV quantum computation as a transformation of the quadrature operators by using the Heisenberg picture. Gaussian operations form an important subset of these, as they act linearly on the quadrature operators in the following sense:
\begin{align}
\hat{\mathbf{q}}_{\textrm{out}}=\mathbf{S}\hat{\mathbf{q}}_{\textrm{in}} + \mathbf{c},\label{eq:symplectic}
\end{align}
where $\hat{\mathbf{q}}=(\hat{x}_{1}\quad\hat{x}_{2}\quad\cdots\quad\hat{x}_{n}\quad\hat{p}_1\quad\hat{p}_2\quad\cdots\quad\hat{p}_{n})^{\textrm{T}}$, and the Heisenberg evolution under the Gaussian operation is represented by $\mathbf{S}$, a $2n\times 2n$ symplectic matrix, and $\mathbf{c}$ a vector of real numbers. Arbitrary operations of this type can be realized via homodyne measurements on CV cluster states that possess an appropriate graph structure \supercite{Asavanant373,Larsen369,PhysRevLett.97.110501}.  The commonly used single-mode Gaussian operations are displacement (contributes a non-zero $c$ vector), phase rotation $\mathbf{R}(\phi)=\begin{pmatrix}
\cos\phi&\sin\phi\\
-\sin\phi&\cos\phi
\end{pmatrix}$, squeezing $\mathbf{S}(\phi)=\begin{pmatrix}
1/\tan\phi&0\\
0&\tan\phi
\end{pmatrix}$, and shear $\mathbf{P}(\phi)=\begin{pmatrix}
1&0\\
2\tan\phi&1
\end{pmatrix}$. To implement universal MBQC, in addition to multi-mode Gaussian operations, non-Gaussian operations are also necessary\supercite{PhysRevLett.82.1784}. Non-Gaussian operations can be realized simply by combining non-Gaussian ancillary states with Gaussian operations\supercite{PhysRevA.64.012310,PhysRevLett.123.200502,PhysRevA.93.022301,PhysRevA.97.022329}.

MBQC using local homodyne measurements on TDM cluster states (Fig.\ \ref{fig:TDMMBQC}a) is equivalent to quantum computation using non-local measurements on two-mode quantum entangled states (Fig.\ \ref{fig:TDMMBQC}b) called two-mode-squeezed states (which become EPR states in the limit of infinite squeezing). In the latter representation, the non-local measurement consists of a 50:50 beamsplitter and two homodyne measurements. The choice of the measurement bases determines which quantum operations are applied. Recall that non-local measurements on two-mode-squeezed states are equivalent to CV gate teleportation, and thus, Fig.\ \ref{fig:TDMMBQC}b can also be interpreted as sequential teleportation.

Figure \ref{fig:TDMMBQC}c shows the experimental setup. TDM consists of using temporal wave packets that propagate along a common light beam and terminate at a fixed detector at different times. The clock frequency of the computation is determined by the time width of the wave packet $\Delta t$. In our current setup, $\Delta t=40$ ns, corresponding to 25 MHz clock frequency. By dynamically changing the measurement bases, we can easily address each mode in a scalable fashion without having to spatially separate them. The homodyne measurement basis is specified by the relative phase between a local oscillator (LO) and an input wave packet. Our programmable homodyne detectors consist of an electro-optic modulator (EOM) and a homemade 2-CH signal generator circuit. By applying the appropriate voltage to the EOM, the circuit can modify the relative phase, thus changing the homodyne measurement basis. To have an independent measurement for each temporal mode, the change in basis must be fast and synchronized. It must also be performed with sufficient precision to ensure high quality quantum gates. Figure \ref{fig:TDMMBQC}d shows an example of the electric signals that our measurement system can generate and use for multi-step MBQC. The technical details of the circuitry can be found in Method.

To evaluate our measurement system and its potential for TDM MBQC, we demonstrate two things: first, we implement various single-step single-mode Gaussian operations in the time domain and demonstrate the programmability of our measurement system. Second, we evaluate its stability for multi-step MBQC of our system via multi-step quantum teleportation. Figure \ref{fig:TDMMBQC}e shows our experimental verification procedure. We consider one mode of the two-mode-squeezed state in the system as the input to time-domain MBQC and the other mode is used as a reference. Quadrature measurement results of the reference mode will be highly correlated with those of the input mode due to entanglement, while the results of each alone will be randomly distributed. Therefore, by calculating correlations between the reference and the output mode, the elements of the symplectic matrix $S$ in Eq.\ \eqref{eq:symplectic} and the nonclassicality of the quantum operations can be verified.

\begin{figure*}[p]
\centering
\includegraphics[width=\textwidth]{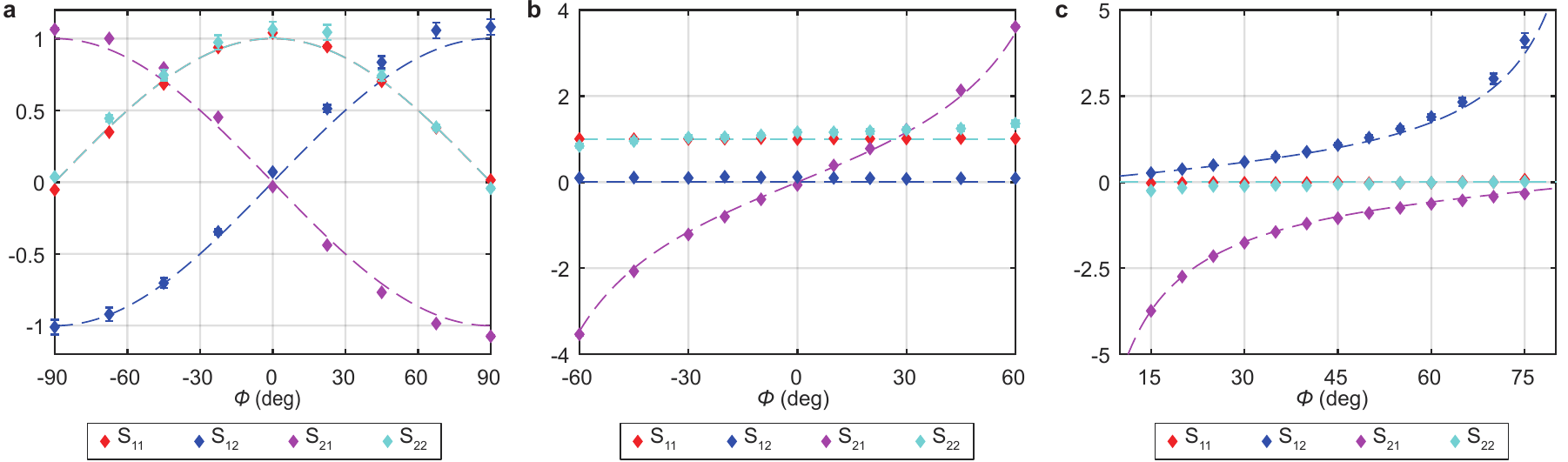}
\caption[]{\textbf{Verification of single-step quantum operations.} The input--output relations are of the form $\begin{pmatrix}
\hat{x}_\textrm{out}\\
\hat{p}_\textrm{out}
\end{pmatrix}=
\begin{pmatrix}
S_{11}&S_{12}\\
S_{21}&S_{22}
\end{pmatrix}
\begin{pmatrix}
\hat{x}_\textrm{in}\\
\hat{p}_\textrm{in}
\end{pmatrix}$. Each subfigure shows plots of $S_{ij}$ of \textbf{a,} phase rotation $\mathbf{R}(\phi)$, \textbf{b,} shear $\mathbf{P}(\phi)$, and \textbf{c,} squeezing with 90$^\circ$ rotation $\mathbf{R}(\pi/2)\mathbf{S}(\phi)$. Each point is obtained using 38,600 correlation measurements (see method). The error bars are also plotted. The dashed lines are ideal theoretical plots.}
\label{fig:Smatrix}
\end{figure*}

In the single-step operations, we implement $\mathbf{R}(\phi)$, $\mathbf{P}(\phi)$, and $\mathbf{R}(\pi/2)\mathbf{S}(\phi)$ gates, for various parameters $\phi$. The parameters $\phi$ are determined by the bases of the homodyne measurement $\theta^\textrm{A}$ and $\theta^\textrm{B}$ (see Method for more details). Note that even for single-step operations, we have to switch between the measurement basis for the reference mode ($\theta_\textrm{ref}$), the operation ($\theta^\textrm{A}$ and $\theta^\textrm{B}$), and the output mode ($\theta_\textrm{out}$) in real time. Figure \ref{fig:Smatrix} shows each element of the $\mathbf{S}$ matrices. In each case, we obtain values for the corresponding symplectic matrices from the experimental data (specifically, the correlations between the quadrature of the reference mode and the output mode). The experimental results are in good agreement with the theoretical predictions. These results indicate that we succeed in the implementations of a variety of single-mode Gaussian operations by programming the measurement bases appropriately. Arbitrary single-mode Gaussian operations can be implemented via combinations of these single-step Gaussian operations.

Next, we verify the nonclassicality of these single-step quantum operations. In this work, we use quantum entanglement as our notion of nonclassicality. Table \ref{tab:result} shows the measurement results of the variances of the nullifiers and verification of quantum entanglement. Prior to MBQC, the reference and output modes are in a two-mode-squeezed state. The infinite squeezing limit of these states are in the zero-eigenspaces of the operators $\hat{x}_\textrm{ref}+\hat{x}_\textrm{in}$ and $\hat{p}_\textrm{ref}-\hat{p}_\textrm{in}$, which are known as nullifiers\supercite{PhysRevA.83.042335}. Hence, they are correlated in momentum and anti-correlated in position. The evolution of the state and corresponding correlations can be efficiently tracked by evolving the nullifiers---the post-evolution nullifiers are denoted by $\hat{\delta}_1$ and $\hat{\delta}_2$. Furthermore, measurement of $\hat{\delta}_1$ and $\hat{\delta}_2$ can be used to verify the nonclassicality of the output. The quantum entanglement between the output mode and the reference mode can be established when the sum of the variances of the nullifiers is below a certain threshold (see Method for derivation). We observe that, for all three operations, the variances of the nullifiers are small (below the vacuum variances), and the successes of the verifications of the quantum entanglement are also shown in Table \ref{tab:result}. Note that the criteria used in this work are sufficient criteria, not necessary and sufficient criteria, meaning that the failures to satisfy the criteria leave the test inconclusive. A few quantum operations that do not satisfy the criteria are those involving large squeezing. See Method for more discussions on this. The non-zero variances are the results of the imperfections in the quantum correlations of the EPR states. We observe that the sum of the variances after the quantum operations are on the order of 1.50 to 1.60 units of vacuum variance for all operations. For single-step operations, theoretically, these values are determined only by the strength of the initial quantum correlations and are independent of the choices of operations.

Results in Fig.\ \ref{fig:Smatrix} and Table \ref{tab:result} indicate the successful implementation of single-step Gaussian operations. Our measurement system can also be used to implement multi-step quantum operations, given that the whole system is sufficiently stable. To demonstrate the stability, we implement $n$-step quantum teleportation (identity operation) for various values of $n$. Then, we calculate the final symplectic matrices $\mathbf{S}$ and the variances of the nullifiers similar to that of Fig.\ \ref{fig:Smatrix} and Table \ref{tab:result}. The experimental results are shown in Fig.\ \ref{fig:multitele}. First, we observe that the $\mathbf{S}$ matrices remain close to an identity matrix for up to $n=100$. On the other hand, the variances of the nullifiers increase as $n$ increases, which is due to the finite squeezing of the system. We plot the theoretical prediction based the variance of the single-step quantum teleportation ($n=1$) and observe that they agree well with the experimental results. This indicates our system is stable, at least up to $n=100$, and there are no other notable imperfections during continuous operation, besides the finite squeezing.

In the verification of the multi-step operation, we select the identity operation for its simplicity in the evaluations. Ideally, the $\mathbf{S}$ matrix remains an identity matrix for all $n$ and the variances of the nullifiers increase linearly with $n$. For arbitrary multi-step operations, the variances of the nullifiers will be highly dependent on how we actually implement the operations, even when the resultant $\mathbf{S}$ matrices are the same\supercite{PhysRevA.90.062324,larsen2020architecture}. This makes the evaluations based on the variances non-trivial for other operations besides repeated identity operation.  Note that for qubit systems, one of the standard procedures for evaluation of the multi-step operations is randomized benchmarking\supercite{PhysRevA.77.012307}, which has also been formulated for qubit-based MBQC\supercite{PhysRevA.94.032303}. For CV systems, however, such experimental procedures are relatively underdeveloped, in addition to the fact that there were no platform capable of large-scale MBQC. We believe that our work will also stimulate developments of an experiment-friendly CV-MBQC version of the randomized benchmarking and might even provide a testbed for it.

In summary, we have developed a programmable homodyne measurement system and used it to demonstrate a prototype of a CV quantum computer based on MBQC multiplexed in time domain with 25 MHz clock frequency. The programmable measurement system made in this work is an essential component for multi-input MBQC, in addition to the previously demonstrated 2D cluster states\supercite{Asavanant373,Larsen369}. The clock frequency is anticipated to reach at least GHz-order with our recently developed THz-bandwidth squeezed light source\supercite{doi:10.1063/1.5142437}. The number of computation steps and the quality of the operations are limited via two factors: the stability of the experimental system and the finite squeezing. Regarding the former, we have explicitly demonstrated that the stability of our system is adequate for at least one hundred steps MBQC and, based on the previous experimental result\supercite{Yokoyama2013}, the system should be stable for at least 10,000 steps of MBQC. Regarding the latter, it is imperative that we increase the squeezing level and add GKP qubits, so that fault-tolerant universal MBQC can be reached\supercite{PhysRevLett.123.200502}. Current state-of-the-art squeezed light sources\supercite{PhysRevLett.117.110801} already admit performance close to the demands set by theoretical predictions for the fault-tolerance threshold\supercite{PhysRevLett.112.120504,PhysRevX.8.021054,PhysRevA.100.010301,PhysRevX.8.021054}. Though technologically challenging, increasing the squeezing level is a relatively straightforward task. Our work is the first working prototype quantum computer based on CV optical systems and it will be interesting to explore the immediate applications of our system in the context of noisy intermediate-scale quantum computation.

% Your references go at the end of the main text, and before the
% figures.  For this document we've used BibTeX, the .bib file
% scibib.bib, and the .bst file Science.bst.  The package scicite.sty
% was included to format the reference numbers according to *Science*
% style.

%BibTeX users: After compilation, comment out the following two lines and paste in
% the generated .bbl file. 

%\bibliography{scibib}

\section*{Method}
\subsection*{Experimental setup, data acquisition, and data analysis}
\begin{figure*}[p]
\centering
\includegraphics{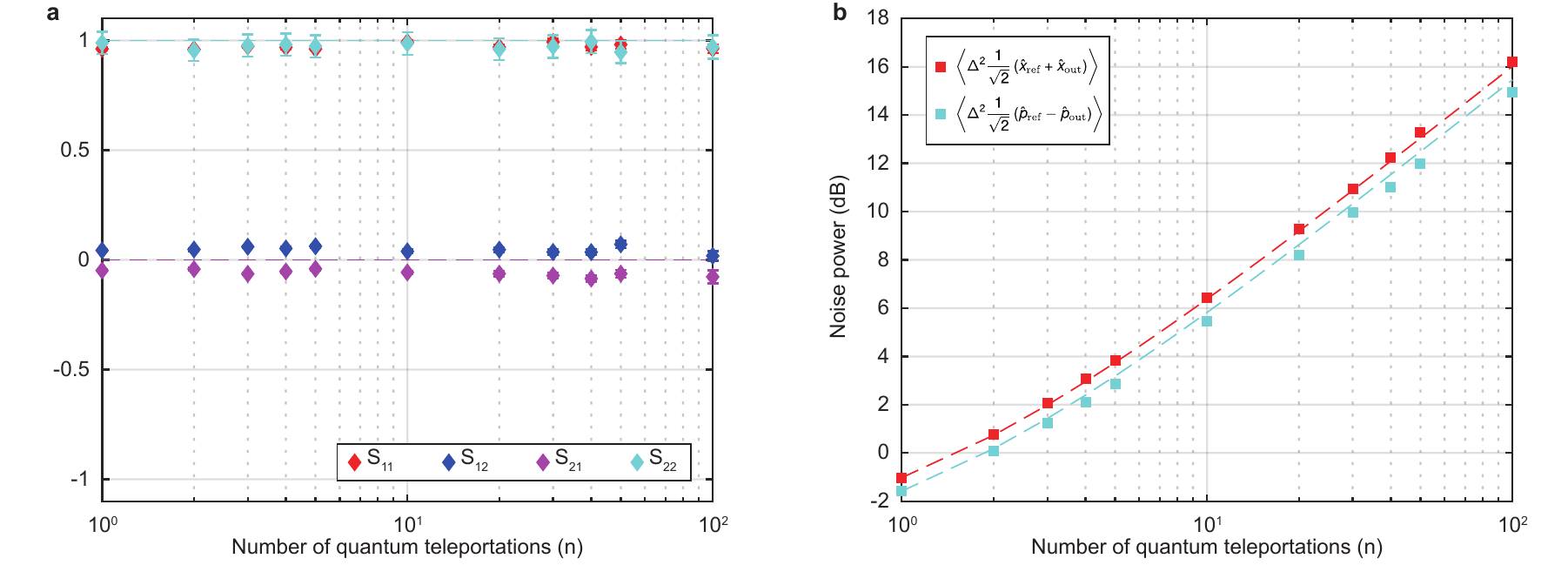}
\caption{\textbf{Multi-step quantum operations.} Identity operations are repeatedly implemented and evaluated. \textbf{a,} Elements of the symplectic matrix $\mathbf{S}$. \textbf{b,} Variances of the nullifiers. 0 dB is the vacuum variance. Each point in both figures are calculated using 38,600 events. Dashed lines: theoretical predictions. Predictions of the variances are based on the results of the single-step identity operation. Errors of the elements of $\mathbf{S}$ matrix are plotted and the errors of the variances are about $\pm0.03$ dB for all points.}
\label{fig:multitele}
\end{figure*}

The cluster state generation system consists of two optical parametric oscillators (OPOs) as the squeezed light sources, 50:50 beam splitters, an optical delay line, and homodyne detectors. The normalized pump amplitudes are set to about 0.7 and the variances of the initial EPR states measured at the homodyne detectors are on average $-4.0$ dB compared to the variance of the vacuum state for both $\hat{x}$ and $\hat{p}$. For detailed specifications of each component, see Ref. \cite{Asavanant373}. The optical delay line is a free-space optical delay line of length 12 m (equivalent to a wave packet size $\Delta t=$ 40 ns). To modify the phase of the homodyne measurement, we use an electro-optical modulator (EOM, Photline, NIR-MPX800-LN-05) and a home-made 2-CH signal generator circuit. The modulators are attached to the local oscillators of the homodyne measurements and their half-wave voltages were measured to be $\sim$4 V. Extended data figure \ref{fig:circuit} shows the schematic of the home-made 2-CH signal generator circuit. This circuit consists of a field-programmable gate array (FPGA; Xilinx, Artix-7), Schmitt triggers (Texas Instruments, SN74LVC1G17), analog adder circuits, and amplifiers (Analog Devices, HMC994APM5E). The FPGA generates synchronized digital signals, necessary for synchronized homodyne basis changes. The signals produced by this FPGA, however, allow switching between only two bases when used as they are. Moreover, the sums of the rise time and the fall time of the signals generated by the FPGA are about 5 ns, which is long compared to the size of the wave packet. Long rise and fall times introduce inaccuracies into the measurement process, thereby degrading the quality of the measurement-based quantum operations. To overcome these issues, we first put the signals from the FPGA into the Schmitt triggers. The sums of the rise and fall times of the outputs of the Schmitt triggers are approximately 700 ps. We then combine outputs of the Schmitt triggers by using the analog adders. The analog adder is composed of inverted amplifiers made from high-speed op-amps (Texas Instruments, OPA855). For each homodyne detector, we combined outputs from seven Schmitt triggers using the analog adders and made a signal generator that can generate 7-bit signals. The signals can be easily changed by programming the FPGA ---the number of the bits directly reflects the phase precision. Note that the phase precision can be simply increased by increasing the number of the Schmitt triggers and is limited only by the circuit noise. With the current number of bits, the phase precision is expected to be between $1^\circ$ to $2^\circ$, which is adequate for current squeezing levels. The sums of the rise and fall times of the output signals of the analog adder are about 2 ns. The signal of the analog adder is then amplified by the analog amplifier before it is sent to the EOM. Bandwidth of the amplifier (28 GHz) is much broader than that of the op-amp used in the analog adder (8 GHz gain bandwidth product). No noticeable overshoot or ringing effects are observed in the output signals and the jitter of the signals are observed to be below 1 ns. Fig.\ \ref{fig:TDMMBQC}d shows an example of the signals that can be generated with this signal generator. Note that $n$-step quantum teleportation in Fig.\ \ref{fig:multitele} involves keeping DC voltages constant for a long time. However, as the current amplifier cannot handle such constant DC signals with long duration, we directly use the signals of the Schmitt triggers for the measurements in Fig.\ \ref{fig:multitele}. In the future, this can be resolved simply by replacing the amplifiers or utilizing EOMs with lower half-wave voltages, so that amplifiers can be removed.

Similar to our recent experiment\supercite{Asavanant373}, we injected phase reference beams into each OPO and used the interference signals between these beams for feedback control and stabilization of the optical system. In that experiment, the phase reference beams and the feedback control were on all the time. Their signals were cut off by electrical low-pass filters as they have low frequency components that are outside the frequency range of the temporal wave packets used. The present experiment introduces additional complications due to the need to change the measurement basis. When the measurement basis is changed, if the phase reference beams also enter the homodyne detectors, the result is high frequency noise that interferes with the measurement results. To prevent this, we use sample \& hold method\supercite{Yokoyama2013}. In the sampling phase, the feedback control and the phase reference beams are on. In the hold phase, they are turned off and the voltage of the electrical components in the feedback systems are held at the values used in the sampling phase. Measurements and changes to the measurement bases are performed during the hold phase. The electrical signals for controlling the timing of the sample \& hold are generated from the same FPGA used in homodyne measurement basis changing, so that all the processes are synchronized. For every measurement, the phase of the homodyne measurements are initially locked in the $\hat{x}$ basis before changing. 

The electrical signals from the homodyne detectors are recorded with an oscilloscope (Tektronix, DP07054). We use a sampling rate of 1 GS/s and a single frame is 10 $\mu$s long, corresponding to 250 wave packets. In each frame, we implement and verify the maximum possible number of operations set by the frame length. For each operation, 38,600 events are used in the estimations of each element of the $S$ matrix. In order to verify the nonclassicality, we also make use of 38,600 events for each linear combination of the quadrature operators for each operation. The quadrature values of the wave packets with index $k$ are calculated by numerically integrating the electrical signals with the mode function $f_{k}(t)$. The mode function we use has the form of
\begin{align}
f_{k}(t)=\begin{cases}
(t-k\Delta t)\exp\left[-\frac{(t-k\Delta t)^{2}}{2\tau^{2}}\right]&\vert t-k\Delta t\vert\leq\frac{\Delta t}{2},\\
0&\vert t-k\Delta t\vert>\frac{\Delta t}{2},
\end{cases}
\end{align} 
with $\tau = 5$ ns. This form is similar to the one in Ref. \cite{Asavanant373}, but with a different parameter to ensure that the wave packet is contained in the time window $\Delta t=40$ ns, even when the effects of the electrical filters are considered (see Ref. \cite{doi:10.1063/1.4962732} for a more detailed discussion). This is necessary especially when we consider switching of the measurement basis. The error bars for all data are calculated using a bootstrapping method.

Feedforward operations (displacements) can be implemented by postprocessing the homodyne measurement data\supercite{PhysRevA.97.032302}. For the explicit formulae for these feedforward operations, see the supplementary information of Ref. \cite{Yokoyama2013}.

When homodyne detectors HD-A and HD-B are set to measure the same basis at the same time bin (blue detectors in Fig.\ \ref{fig:TDMMBQC}d), the result is a logical measurement of the input encoded at that time step (rather than the implementation of a measurement-based gate, which occurs when the homodyne bases at A and B differ)\supercite{PhysRevA.97.032302}. We perform this measurement on both the reference mode, and the output mode.
%The measurements of the reference mode and the output mode (blue detectors in Fig.\ \ref{fig:TDMMBQC}d) is obtained by selecting the same measurement basis for HD-A and HD-B and post processing the measurement results\supercite{PhysRevA.97.032302}. 
If we let $\hat{x}(\theta)=(\cos\theta)\hat{x}+(\sin\theta)\hat{p}$, then the quadratures of the reference mode and the input mode are
\begin{align}
\hat{x}_\textrm{ref}(\theta)&=\frac{1}{\sqrt{2}}\left[\hat{x}^\textrm{A}(\theta)-\hat{x}^\textrm{B}(\theta)\right],\\
\hat{x}_\textrm{out}(\theta)&=\frac{1}{\sqrt{2}}\left[\hat{x}^\textrm{A}(\theta)+\hat{x}^\textrm{B}(\theta)\right].
\end{align}

\subsection*{Quantum teleportation, Gaussian operations, and actual noise}
In the limit of large squeezing, selecting measurement bases $\hat{x}^{\textrm{A}}(\theta^\textrm{A})=(\cos\theta^\textrm{A})\hat{x}^{\textrm{A}}+(\sin\theta^\textrm{A})\hat{p}^{\textrm{A}}$ and $\hat{x}^{\textrm{B}}(\theta^\textrm{B})=(\cos\theta^\textrm{B})\hat{x}^{\textrm{B}}+(\sin\theta^\textrm{B})\hat{p}^{\textrm{B}}$ implements the following Heisenberg-picture map\supercite{PhysRevA.97.032302}
\begin{align}
\hat{\mathbf{q}}_{\textrm{out}}=\mathbf{V}(\theta^\textrm{A},\theta^\textrm{B})\hat{\mathbf{q}}_{\textrm{in}},\label{eq:quantumtele}
\end{align}
where $\hat{\mathbf{q}}=(\hat{x}\quad\hat{p})^\textrm{T}$ and
\begin{align}
\mathbf{V}(\theta^\textrm{A},\theta^\textrm{B})=\mathbf{R}(\theta_{+}-\pi/2)\mathbf{S}(\theta_{-})\mathbf{R}(\theta_{+}),
\end{align}
with $\theta_{\pm}=(\theta^\textrm{B}\pm\theta^\textrm{A})/2$, $\mathbf{R}(\cdot)$ and $\mathbf{S}(\cdot)$ defined the same way as in the main text, and $\theta^\textrm{A}\neq\theta^\textrm{B}$. The identity operation, i.e. quantum teleportation, corresponds to the case where $\theta^\textrm{A}=0$ and $\theta^\textrm{B}=\pi/2$. Phase rotations $\mathbf{R}(\phi)$ can be implemented by setting $\theta^\textrm{A}=\phi/2$ and $\theta^\textrm{B}=\phi/2+\pi/2$. Squeezing operations (with an additional $\pi/2$ phase rotation) $\mathbf{R}(\pi/2)\mathbf{S}(\phi)$ correspond to the case where $\theta^\textrm{A}=\phi$ and $\theta^\textrm{B}=-\phi$. Shear operations $\mathbf{P}(\phi)$ correspond to the case where $\theta^\textrm{A}=0$ and $\theta^\textrm{B}=\pi/2-\phi$. 

In the case of a finitely squeezed resource and in the absence of optical losses, Eq.\ \eqref{eq:quantumtele} must be modified:
\begin{align}
\hat{\mathbf{q}}_{\textrm{out}}=\mathbf{V}(\theta^\textrm{A},\theta^\textrm{B})\hat{\mathbf{q}}_{\textrm{in}}+
\begin{pmatrix}
\hat{x}_{\textrm{anc},1}e^{-r_{x}}\\
\hat{p}_{\textrm{anc},2}e^{-r_{p}}
\end{pmatrix}.
\end{align}
The above includes a second term that captures the noise due to finite squeezing. This contains $\hat{x}_{\textrm{anc},1}$ and $\hat{p}_{\textrm{anc},2}$, which are quadrature operators of ancillary vacuum states with squeezing parameters $r_{x}$ and $r_{p}$. The unitary part $\mathbf{V}(\theta^\textrm{A},\theta^\textrm{B})$ is identical to the infinite squeezing case. Since the unitary part is not affected by finite squeezing effects, we can verify it separately.  Note that in the actual experiment, there are optical losses in various components, which makes the actual relations more complicated. However, the fact that $\mathbf{V}(\theta^\textrm{A},\theta^\textrm{B})$ depends mainly on the choice of measurement basis and the noise of the operations are determined by the amount of the squeezing in the system does not change.

\subsection*{Estimation of $\mathbf{S}$ matrix}
The general form of a single-mode Gaussian operation (without displacements) including Gaussian noise is
\begin{align}
\begin{pmatrix}
\hat{x}_\textrm{out}\\
\hat{p}_\textrm{out}
\end{pmatrix}=\begin{pmatrix}
S_{11}&S_{12}\\
S_{21}&S_{22}
\end{pmatrix}
\begin{pmatrix}
\hat{x}_\textrm{in}\\
\hat{p}_\textrm{in}
\end{pmatrix}+\hat{\mathbf{N}},\label{eq:suppS}
\end{align}
where $\hat{\mathbf{N}}$ is a noise term and $S$ is no longer restricted to a symplectic matrix. We remove this assumption because it no longer holds when noise is involved. For example, if the input mode were entirely lost, then $S=0$ and the contributions to the output quadratures come entirely from noise. As the quadrature values of the input mode and the output mode cannot be simultaneously measured, we use a method that does not involve such measurement.

In this work, we measure the values of each element of $S$ by using quantum entanglement. EPR correlation between the input and reference modes is given by
\begin{align}
\begin{pmatrix}
\hat{x}_\textrm{ref}\\
\hat{p}_\textrm{ref} 
\end{pmatrix}=
\begin{pmatrix}
-\hat{x}_\textrm{in}\\
\hat{p}_\textrm{in}
\end{pmatrix}+\hat{\mathbf{N}}^\prime,
\end{align}
where $\hat{\mathbf{N}}^\prime$ is a noise term due to the finite squeezing which satisfies $\mean{\hat{\mathbf{N}}^\prime}=\mathbf{0}$. Thus, by measuring the reference modes, we can infer the quadrature values of the input modes. Using the fact that $\mean{\hat{x}_\textrm{ref}\hat{p}_\textrm{in}}=\mean{\hat{x}_\textrm{in}\hat{p}_\textrm{ref}}=0$ for the EPR states, we can directly obtain the elements of $S$ from the quadrature values by calculating
\begin{align}
S_{ij}=\frac{\mean{\hat{\xi}_\textrm{out}^{(i)}\hat{\xi}_\textrm{ref}^{(j)}}}{\mean{\hat{\xi}_\textrm{in}^{(j)}\hat{\xi}_\textrm{ref}^{(j)}}},\label{eq:Sestimation}
\end{align}
where $\hat{\xi}^{(1)}=\hat{x}$ and $\hat{\xi}^{(2)}=\hat{p}$.

\subsection*{Verification of nonclassicality and criteria for quantum inseparability}
The van Loock-Furusawa criteria\supercite{PhysRevA.67.052315} are often used to verify genuine multi-partite inseparability in CV cluster state experiments\supercite{Asavanant373,Larsen369,doi:10.1063/1.4962732,Yokoyama2013,PhysRevLett.112.120505}. These criteria are formulated in terms of variances of observables that consist of linear combinations of only either $\hat{x}$ or $\hat{p}$ across different modes. In the current experiment, however, the relationships between the output mode and the input mode as indicated by Eq.\ \eqref{eq:symplectic} are not limited to relationships containing only $\hat{x}$ or $\hat{p}$.

Here we use a more general form of inseparability criteria that is applicable to our experiment\supercite{Ukai00}. First, let us consider two subsystems A and B. Then, let us consider operators $\hat{\delta}_1$, $\hat{\delta}_2$ which have the following form
\begin{align}
\hat{\delta}_1&=\hat{\zeta}^{\textrm{A}}+\hat{\zeta}^{\textrm{B}},\\
\hat{\delta}_2&=\hat{\xi}^{\textrm{A}}+\hat{\xi}^{\textrm{B}},
\end{align}
where the superscripts A, B are used to denote which subsystem the operators belongs to, and $\hat{\zeta}$ and $\hat{\xi}$ are arbitrary linear combinations of both $\hat{x}$ and $\hat{p}$ type operators. Then, similar to the proof of van Loock-Furusawa criteria\supercite{PhysRevA.67.052315}, we can show the following: if the subsystems are separable, i.e. the density matrix is in the form $\hat{\rho}=\sum_{i}\eta_{i}\hat{\rho}^{\textrm{A}}_{i}\otimes\hat{\rho}^{\textrm{B}}_{i}$ with $\sum_{i}\eta_{i}=1$, then, the following relation holds:
\begin{align}
\left\langle\Delta^2\hat{\delta}_1\right\rangle+\left\langle\Delta^2\hat{\delta}_2\right\rangle\geq\vert\langle[\hat{\zeta}^{\textrm{A}},\hat{\xi}^{\textrm{A}}]\rangle\vert+\vert\langle[\hat{\zeta}^{\textrm{B}},\hat{\xi}^{\textrm{B}}]\rangle\vert,\label{eq:inseparabilitycriteria}
\end{align}
where $\langle\cdot\rangle$ denotes the mean and $\langle\Delta^2\cdot\rangle$ denotes the variance. For the linear combination of the quadrature operators, the right hand side of Eq. \eqref{eq:inseparabilitycriteria} evaluates to a constant, thus does not depend on $\hat{\rho}$. Therefore, violation of the above inequalities provide sufficient criteria for two-mode quantum inseparability. Note that this derivation does not tell us the best choices of linear combinations to use for any given state. It might seem natural to use the nullifiers since their variances (i.e. left side of Eq.\ \eqref{eq:inseparabilitycriteria}) tend to 0 in the infinite squeezing limit. When we consider the finite squeezing case, however, this is not necessarily an optimal choice. We leave the exploration of such a possibility to future work and use the nullifiers for our verification procedure.

From Eq.\ \eqref{eq:inseparabilitycriteria} it is clear that the thresholds for establishing quantum inseparability depend on the  quantum operations we perform. As an example, let us consider the variance of the nullifiers as they evolve under squeezing operations: $(\sin\phi)\hat{p}_{\textrm{out}}-(\cos\phi)\hat{x}_{\textrm{ref}}$ and $(\cos\phi)\hat{x}_{\textrm{out}}-(\sin\phi)\hat{p}_{\textrm{ref}}$. Theoretically, the variances of these operators are determined by the squeezing level in the system and have the same values for all $\phi$ (i.e. same for all squeezing parameters of the squeezing operations). On the other hand, the right hand side of Eq.\ \eqref{eq:inseparabilitycriteria} is $\hbar\vert\sin2\phi\vert$, which means that the larger the squeezing factor is (i.e. the closer $\phi$ is to 0 or $\pi/2$) the more severe the threshold becomes. Therefore, if we want to implement a gate that involves high squeezing and still observe quantum entanglement using this criteria, we would expect that the initial resource state must be highly squeezed.

\subsection*{Theoretical prediction of the variances in Fig.\ \ref{fig:multitele}b}
When we implement $n$-step identity operations, theoretically, the variances will increase linearly with the number of steps. If we measure the nullifiers after a single-step, then the variances become
\begin{align}
\mean{\Delta^2\left[\frac{1}{\sqrt{2}}\left(\hat{x}_\textrm{out}+\hat{x}_\textrm{ref}\right)\right]}&=2\times\frac{\hbar}{2}e^{-2r_x},\\
\mean{\Delta^2\left[\frac{1}{\sqrt{2}}\left(\hat{p}_\textrm{out}-\hat{p}_\textrm{ref}\right)\right]}&=2\times\frac{\hbar}{2}e^{-2r_p},
\end{align}
where we assume that the two-mode-squeezed states used as both the input and the quantum teleportation(s) have the same level of squeezing. Hence, we can infer increases in variances after $n$ steps and draw a theoretical plot. After $n$ steps, the variances of the nullifiers become
\begin{align}
\mean{\Delta^2\left[\frac{1}{\sqrt{2}}\left(\hat{x}_\textrm{out}+\hat{x}_\textrm{ref}\right)\right]}_{n}&=(n+1)\times\frac{\hbar}{2}e^{-2r_x},\\
\mean{\Delta^2\left[\frac{1}{\sqrt{2}}\left(\hat{p}_\textrm{out}-\hat{p}_\textrm{ref}\right)\right]}_{n}&=(n+1)\times\frac{\hbar}{2}e^{-2r_p}.
\end{align}
The above equations are used in the theoretical plots of Fig.\ \ref{fig:multitele}b. 
%% reference %%
%\nocite{*}

\printbibliography

\section*{Acknowledgments}
\subsection*{Funding}
This work was partly supported by JSPS KAKENHI (Grant No.\ 18H05207), the Core Research for Evolutional
Science and Technology (CREST) (Grant No.\ JPMJCR15N5) of the Japan Science and Technology Agency (JST), UTokyo Foundation, donations from Nichia Corporation, and the Australian Research Council Centre of Excellence for Quantum Computation and Communication Technology (Project No. CE170100012). W.A. acknowledges financial support from the Japan Society for the Promotion of Science (JSPS). B.C. acknowledges financial support from Program of Excellence in Photon Science (XPS). T.N. acknowledges financial support from Forefront Physics and Mathematics Program to Drive Transformation (FoPM). R. N. A. is supported by National Science Foundation grant PHY-1630114.

\subsection*{Author contributions}
W.A. and B.C. conceive the project and the experiment was supervised by S.Y., M.E., J.Y., H.Y., and A.F. R.N.A., N.C.M., and W.A. work on the theory for the experiment. B.C., W.A. and T.N. build the optical part of the setup. B.C., T.E., W.A. and M.E. design, make, and test the electrical circuits for homodyne measurement basis switching. B.C. and T.E. write the code to program FPGA in quantum operations. W.A. and B.C. carry out the experiment and acquire the experimental data. W.A. perform data analysis, visualization of experimental data, and theoretical predictions of the experimental results. W.A. write the manuscript with assistance from all the co-authors.

\subsection*{Competing financial interests}
The authors declare no competing financial interests.

\subsection*{Data and materials availability}
Data that supporting the finding are available upon reasonable request. Any correspondence and request should be addressed to A. F.

\begin{table}[p]
\centering
\caption{\textbf{Variances and entanglement threshold of various quantum operations.}}
\resizebox{\textwidth}{!}{

%\rowcolors{1}{}{gray!25}
\begin{tabular}{cccccccccc}

%table heading
\hline
\rowcolor{gray!50}&&\multicolumn{2}{c}{Nullifiers}&&\multicolumn{3}{c}{Variances of the nullifiers}&&\\
\multirow{-2}{*}{\cellcolor{gray!50}Operation}&\multirow{-2}{*}{\cellcolor{gray!50}Relation}&\cellcolor{gray!50}$\hat{\delta}_1$&\cellcolor{gray!50}$\hat{\delta}_2$&\multirow{-2}{*}{\cellcolor{gray!50}$\phi\,(^\circ)$}&\cellcolor{gray!50}$\hat{\delta}_1$&\cellcolor{gray!50}$\hat{\delta}_2$&\cellcolor{gray!50}Sum&\multirow{-2}{*}{\cellcolor{gray!50}Threshold}&\multirow{-2}{*}{\cellcolor{gray!50}\Checkmark}\\
\hline

%Phase rotation
\cellcolor{white}&\cellcolor{white}&\cellcolor{white}&\cellcolor{white}&$0$&$0.80\pm0.01$&$0.83\pm0.01$&$1.63\pm0.02$&\cellcolor{white}&\Checkmark\\

\cellcolor{white}&\cellcolor{white}&\cellcolor{white}&\cellcolor{white}&\cellcolor{gray!25}$\cellcolor{gray!25}22.5$&\cellcolor{gray!25}$0.81\pm0.01$&\cellcolor{gray!25}$0.68\pm0.01$&\cellcolor{gray!25}$1.49\pm0.02$&\cellcolor{white}&\cellcolor{gray!25}\Checkmark\\

\cellcolor{white}&\cellcolor{white}&\cellcolor{white}&\cellcolor{white}&$-22.5$&$0.81\pm0.01$&$0.68\pm0.01$&$1.49\pm0.02$&\cellcolor{white}&\cellcolor{white}\Checkmark\\

\cellcolor{white}&\cellcolor{white}&\cellcolor{white}&\cellcolor{white}&\cellcolor{gray!25}$45$&\cellcolor{gray!25}$0.80\pm0.01$&\cellcolor{gray!25}$0.68\pm0.01$&\cellcolor{gray!25}$1.48\pm0.02$&\cellcolor{white}&\cellcolor{gray!25}\Checkmark\\

\cellcolor{white}&\cellcolor{white}&\cellcolor{white}&\cellcolor{white}&$-45$&$0.80\pm0.01$&$0.69\pm0.01$&$1.49\pm0.02$&\cellcolor{white}&\cellcolor{white}\Checkmark\\

\cellcolor{white}&\cellcolor{white}&\cellcolor{white}&\cellcolor{white}&\cellcolor{gray!25}$67.5$&\cellcolor{gray!25}$0.78\pm0.01$&\cellcolor{gray!25}$0.72\pm0.01$&\cellcolor{gray!25}$1.50\pm0.02$&\cellcolor{white}&\cellcolor{gray!25}\Checkmark\\

\cellcolor{white}&\cellcolor{white}&\cellcolor{white}&\cellcolor{white}&$-67.5$&$0.81\pm0.01$&$0.70\pm0.01$&$1.51\pm0.02$&\cellcolor{white}&\Checkmark\\

\cellcolor{white}&\cellcolor{white}&\cellcolor{white}&\cellcolor{white}&\cellcolor{gray!25}$90$&\cellcolor{gray!25}$0.77\pm0.01$&\cellcolor{gray!25}$0.73\pm0.01$&\cellcolor{gray!25}$1.50\pm0.02$&\cellcolor{white}&\cellcolor{gray!25}\Checkmark\\

\multirow{-9}{*}{\cellcolor{white}Phase rotation}&\multirow{-9}{*}{\cellcolor{white}$\begin{pmatrix}
\hat{x}_{\textrm{out}}\\
\hat{p}_{\textrm{out}}
\end{pmatrix}=
\begin{pmatrix}
-\cos\phi&\sin\phi\\
\sin\phi&\cos\phi
\end{pmatrix}
\begin{pmatrix}
\hat{x}_{\textrm{ref}}\\
\hat{p}_{\textrm{ref}}
\end{pmatrix}$}&\multirow{-9}{*}{\cellcolor{white}$\frac{1}{\sqrt{2}}[\hat{x}_{\textrm{out}}(-\phi/2)+\hat{x}_{\textrm{ref}}(-\phi/2)]$}&\multirow{-9}{*}{\cellcolor{white}$\frac{1}{\sqrt{2}}[\hat{p}_{\textrm{out}}(-\phi/2)-\hat{p}_{\textrm{ref}}(-\phi/2)]$}&$-90$&$0.77\pm0.01$&$0.72\pm0.01$&$1.49\pm0.02$&\multirow{-9}{*}{\cellcolor{white}2.00}&\Checkmark\\

%Squeezing
\cellcolor{gray!25}&\cellcolor{gray!25}&\cellcolor{gray!25}&\cellcolor{gray!25}&\cellcolor{gray!25}$15$&\cellcolor{gray!25}$0.81\pm0.01$&\cellcolor{gray!25}$0.78\pm0.01$&\cellcolor{gray!25}$1.59\pm0.02$&\cellcolor{gray!25}1.00&\cellcolor{gray!25}\\

\cellcolor{gray!25}&\cellcolor{gray!25}&\cellcolor{gray!25}&\cellcolor{gray!25}&$20$&$0.82\pm0.01$&$0.81\pm0.01$&$1.63\pm0.02$&1.29&\\

\cellcolor{gray!25}&\cellcolor{gray!25}&\cellcolor{gray!25}&\cellcolor{gray!25}&\cellcolor{gray!25}$25$&\cellcolor{gray!25}$0.82\pm0.01$&\cellcolor{gray!25}$0.78\pm0.01$&\cellcolor{gray!25}$1.60\pm0.02$&\cellcolor{gray!25}1.53&\cellcolor{gray!25}\\

\cellcolor{gray!25}&\cellcolor{gray!25}&\cellcolor{gray!25}&\cellcolor{gray!25}&$30$&$0.83\pm0.01$&$0.77\pm0.01$&$1.60\pm0.02$&1.73&\Checkmark\\

\cellcolor{gray!25}&\cellcolor{gray!25}&\cellcolor{gray!25}&\cellcolor{gray!25}&\cellcolor{gray!25}$35$&\cellcolor{gray!25}$0.84\pm0.01$&\cellcolor{gray!25}$0.80\pm0.01$&\cellcolor{gray!25}$1.64\pm0.02$&\cellcolor{gray!25}1.87&\cellcolor{gray!25}\Checkmark\\

\cellcolor{gray!25}&\cellcolor{gray!25}&\cellcolor{gray!25}&\cellcolor{gray!25}&$40$&$0.81\pm0.01$&$0.77\pm0.01$&$1.58\pm0.02$&1.96&\Checkmark\\

\cellcolor{gray!25}&\cellcolor{gray!25}&\cellcolor{gray!25}&\cellcolor{gray!25}&\cellcolor{gray!25}$45$&\cellcolor{gray!25}$0.80\pm0.01$&\cellcolor{gray!25}$0.79\pm0.01$&\cellcolor{gray!25}$1.59\pm0.02$&\cellcolor{gray!25}2.00&\cellcolor{gray!25}\Checkmark\\

\cellcolor{gray!25}&\cellcolor{gray!25}&\cellcolor{gray!25}&\cellcolor{gray!25}&$50$&$0.82\pm0.01$&$0.77\pm0.01$&$1.59\pm0.02$&1.96&\Checkmark\\

\cellcolor{gray!25}&\cellcolor{gray!25}&\cellcolor{gray!25}&\cellcolor{gray!25}&\cellcolor{gray!25}$55$&\cellcolor{gray!25}$0.81\pm0.01$&\cellcolor{gray!25}$0.77\pm0.01$&\cellcolor{gray!25}$1.58\pm0.02$&\cellcolor{gray!25}1.87&\cellcolor{gray!25}\Checkmark\\

\cellcolor{gray!25}&\cellcolor{gray!25}&\cellcolor{gray!25}&\cellcolor{gray!25}&$60$&$0.80\pm0.01$&$0.77\pm0.01$&$1.57\pm0.02$&1.73&\Checkmark\\

\cellcolor{gray!25}&\cellcolor{gray!25}&\cellcolor{gray!25}&\cellcolor{gray!25}&\cellcolor{gray!25}$65$&\cellcolor{gray!25}$0.84\pm0.01$&\cellcolor{gray!25}$0.78\pm0.01$&\cellcolor{gray!25}$1.62\pm0.02$&\cellcolor{gray!25}1.53&\cellcolor{gray!25}\\

\cellcolor{gray!25}&\cellcolor{gray!25}&\cellcolor{gray!25}&\cellcolor{gray!25}&$70$&$0.84\pm0.01$&$0.76\pm0.01$&$1.60\pm0.02$&1.29&\\

\multirow{-13}{*}{\cellcolor{gray!25}Squeezing}&\multirow{-13}{*}{\cellcolor{gray!25}$\begin{pmatrix}
\hat{x}_{\textrm{out}}\\
\hat{p}_{\textrm{out}}
\end{pmatrix}=\begin{pmatrix}
0&\tan\phi\\
\frac{1}{\tan\phi}&0
\end{pmatrix}
\begin{pmatrix}
\hat{x}_{\textrm{ref}}\\
\hat{p}_{\textrm{ref}}
\end{pmatrix}$}&\multirow{-13}{*}{\cellcolor{gray!25}$(\cos\phi)\hat{x}_{\textrm{out}}-(\sin\phi)\hat{p}_{\textrm{ref}}$}&\multirow{-13}{*}{\cellcolor{gray!25}$(\sin\phi)\hat{p}_{\textrm{out}}-(\cos\phi)\hat{x}_{\textrm{ref}}$}&\cellcolor{gray!25}$75$&\cellcolor{gray!25}$0.81\pm0.01$&\cellcolor{gray!25}$0.76\pm0.01$&\cellcolor{gray!25}$1.57\pm0.02$&\cellcolor{gray!25}1.00&\cellcolor{gray!25}\\

%shear
\multirow{11}{*}{Shear}&\multirow{11}{*}{$\begin{pmatrix}
\hat{x}_{\textrm{out}}\\
\hat{p}_{\textrm{out}}
\end{pmatrix}=
\begin{pmatrix}
-1&0\\
-2\tan\phi&1
\end{pmatrix}
\begin{pmatrix}
\hat{x}_{\textrm{ref}}\\
\hat{p}_{\textrm{ref}}
\end{pmatrix}$}&\multirow{11}{*}{$\frac{1}{\sqrt{2}}(\hat{x}_{\textrm{out}}+\hat{x}_{\textrm{ref}})$}&\multirow{11}{*}{$\frac{1}{\sqrt{2}}\left[\hat{p}_{\textrm{out}}(\phi)-\hat{p}_{\textrm{ref}}(\phi)\right]$}&$0$&$0.82\pm0.01$&$0.89\pm0.01$&$1.71\pm0.02$&2.00&\Checkmark\\

&&&&\cellcolor{gray!25}$10$&\cellcolor{gray!25}$0.82\pm0.01$&\cellcolor{gray!25}$0.70\pm0.01$&\cellcolor{gray!25}$1.54\pm0.02$&\cellcolor{gray!25}1.96&\cellcolor{gray!25}\Checkmark\\

&&&&$-10$&$0.85\pm0.01$&$0.72\pm0.01$&$1.57\pm0.02$&1.96&\Checkmark\\

&&&&\cellcolor{gray!25}$20$&\cellcolor{gray!25}$0.82\pm0.01$&\cellcolor{gray!25}$0.71\pm0.01$&\cellcolor{gray!25}$1.53\pm0.02$&\cellcolor{gray!25}1.87&\cellcolor{gray!25}\Checkmark\\

&&&&$-20$&$0.85\pm0.01$&$0.72\pm0.01$&$1.57\pm0.02$&1.87&\Checkmark\\

&&&&\cellcolor{gray!25}$30$&\cellcolor{gray!25}$0.85\pm0.01$&\cellcolor{gray!25}$0.74\pm0.01$&\cellcolor{gray!25}$1.59\pm0.02$&\cellcolor{gray!25}1.73&\cellcolor{gray!25}\Checkmark\\

&&&&$-30$&$0.83\pm0.01$&$0.73\pm0.01$&$1.56\pm0.02$&1.73&\Checkmark\\

&&&&\cellcolor{gray!25}$45$&\cellcolor{gray!25}$0.83\pm0.01$&\cellcolor{gray!25}$0.75\pm0.01$&\cellcolor{gray!25}$1.58\pm0.02$&\cellcolor{gray!25}1.41&\cellcolor{gray!25}\\

&&&&$-45$&$0.83\pm0.01$&$0.75\pm0.01$&$1.58\pm0.02$&1.41&\\

&&&&\cellcolor{gray!25}$60$&\cellcolor{gray!25}$0.81\pm0.01$&\cellcolor{gray!25}$0.77\pm0.01$&\cellcolor{gray!25}$1.58\pm0.02$&\cellcolor{gray!25}1.00&\cellcolor{gray!25}\\

&&&&$-60$&$0.83\pm0.01$&$0.78\pm0.01$&$1.61\pm0.02$&1.00&\\

\hline
\end{tabular}}
\begin{flushright}
{\footnotesize Note that $\hat{x}(\phi)=(\cos\phi)\hat{x}+(\sin\phi)\hat{p}$ and $\hat{p}(\phi)=(\cos\phi)\hat{p}-(\sin\phi)\hat{x}$.\\ The checkmarks \Checkmark indicate whether the sums of the variances are below the inseparability threshold or not.\\ Variances are shown in vacuum variances unit (i.e. multiples of $\hbar/2$).}
\end{flushright}
\label{tab:result}
\end{table}

\setcounter{figure}{0}
\renewcommand{\figurename}{Extended data figure}
\begin{figure*}[p]
\centering
\includegraphics{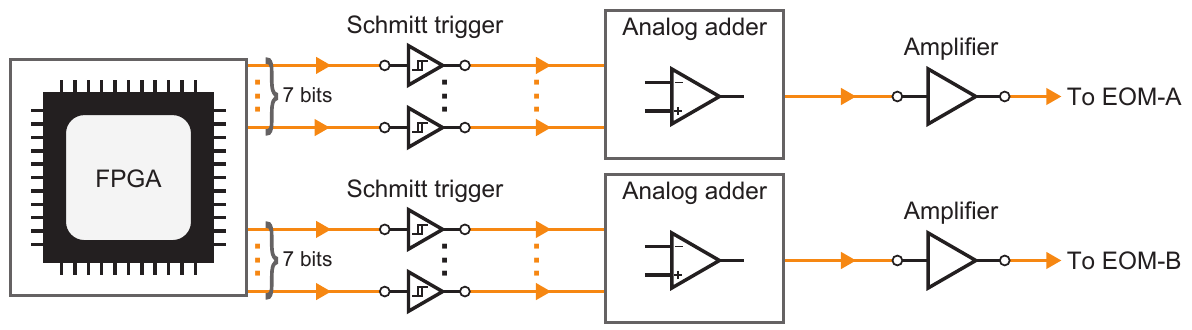}
\caption{\textbf{Schematic diagram of our home-made 2-CH signal generator circuit.} }
\label{fig:circuit}
\end{figure*}

%\begin{figure*}[p]
%\centering
%\includegraphics{figure/waveform.pdf}
%\caption{\textbf{Example of the waveforms that can be generated after the analog adder} }
%\label{fig:waveform}
%\end{figure*}

%Here you should list the contents of your Supplementary Materials -- below is an example. 
%You should include a list of Supplementary figures, Tables, and any references that appear only in the SM. 
%Note that the reference numbering continues from the main text to the SM.
% In the example below, Refs. 4-10 were cited only in the SM.     

% For your review copy (i.e., the file you initially send in for
% evaluation), you can use the {figure} environment and the
% \includegraphics command to stream your figures into the text, placing
% all figures at the end.  For the final, revised manuscript for
% acceptance and production, however, PostScript or other graphics
% should not be streamed into your compliled file.  Instead, set
% captions as simple paragraphs (with a \noindent tag), setting them
% off from the rest of the text with a \clearpage as shown  below, and
% submit figures as separate files according to the Art Department's
% instructions.

\end{document}